\RequirePackage{lineno}
\documentclass[aps,twocolumn,superscriptaddress,preprintnumbers,amsmath,amssymb]{revtex4}
\usepackage{graphicx}
\usepackage{epsfig}
\usepackage{array} 
\usepackage{dcolumn}
\usepackage{bm}
\usepackage{amsmath}
\usepackage{color}
\usepackage[colorlinks,linkcolor=red,anchorcolor=green,citecolor=blue]{hyperref}
\usepackage{multirow}
\usepackage{relsize}
\usepackage{verbatim}
\usepackage{orcidlink}

\newcommand{\BR}{{\cal B}}

\newcommand{\gev}{\rm GeV}

\newcommand{\beq}{\begin{equation}}
\newcommand{\eeq}{\end{equation}}
\newcommand{\bitm}{\begin{itemize}}
\newcommand{\eitm}{\end{itemize}}

\begin{document}

\title{\quad\\[1.0cm] Search for $e^+ e^- \to \gamma\chi_{bJ}$ ($J$ = 0, 1, 2) near $\sqrt{s} = 10.746$ GeV at Belle II}


  \author{M.~Abumusabh\,\orcidlink{0009-0004-1031-5425}} 
  \author{I.~Adachi\,\orcidlink{0000-0003-2287-0173}} 
  \author{L.~Aggarwal\,\orcidlink{0000-0002-0909-7537}} 
  \author{H.~Ahmed\,\orcidlink{0000-0003-3976-7498}} 
  \author{Y.~Ahn\,\orcidlink{0000-0001-6820-0576}} 
  \author{H.~Aihara\,\orcidlink{0000-0002-1907-5964}} 
  \author{N.~Akopov\,\orcidlink{0000-0002-4425-2096}} 
  \author{S.~Alghamdi\,\orcidlink{0000-0001-7609-112X}} 
  \author{M.~Alhakami\,\orcidlink{0000-0002-2234-8628}} 
  \author{A.~Aloisio\,\orcidlink{0000-0002-3883-6693}} 
% ...
\collaboration{The Belle II Collaboration}

\begin{abstract}

We search for the $e^+ e^- \to \gamma \chi_{bJ}$ ($J$ = 0, 1, 2) processes at center-of-mass energies $\sqrt{s}$ = 10.653, 10.701, 10.746, and 10.804 GeV. These data were collected with the Belle II detector at the SuperKEKB collider and correspond to 3.5, 1.6, 9.8, and 4.7 fb$^{-1}$ of integrated luminosity, respectively. We set upper limits at the 90\% confidence level on the Born cross sections for $e^+ e^- \to \gamma \chi_{bJ}$ at each center-of-mass energy $\sqrt{s}$ near 10.746 GeV.
The upper limits at 90\% confidence level on the Born cross sections for $e^+ e^- \to \gamma \chi_{b1}$ are significantly smaller than the corresponding measured values for $e^+e^-\to\omega\chi_{b1}$ and $e^+e^-\to\pi^+\pi^-\Upsilon(2S)$ at $\sqrt{s}$ = 10.746 GeV.

\end{abstract}

\maketitle

In 2019, a resonance with mass $(10752.7^{+5.9}_{-6.0})$ MeV/$c^2$ and width $(35.5^{+18.0}_{-11.8})$ MeV was observed in the $e^+ e^-\to \Upsilon(nS)\pi^+\pi^-$ ($n$ = 1, 2, 3) process by Belle, which is denoted $\Upsilon(10753)$~\cite{220}. The $\Upsilon(10753)$ has been interpreted as a conventional bottomonium resonance~\cite{074007,034036,59,014020,014036,357,04049,135340,11915,103845,Ni:2025gvx}, a hybrid~\cite{034019,1}, or a tetraquark state~\cite{135217,074507,11475,123102,381,Zhao:2025kno}.
To understand the nature of this new resonance, further studies of $\Upsilon(10753)$ decays are needed.

Belle II has performed a number of studies using data collected at $e^+e^-$ center-of-mass energy $\sqrt{s}$ near 10.746 GeV. The observation of $e^+e^-\to\omega\chi_{bJ}$ ($J$ = 1, 2) at $\sqrt{s}=$ 10.746 GeV and evidence for $e^+e^-\to\omega\chi_{bJ}$ ($J$ = 1, 2) at $\sqrt{s}=$ 10.804 GeV were reported using the full data sample of 19.6 fb$^{-1}$ at $\sqrt{s}$ near 10.75 GeV~\cite{Belle-II:2022xdi}, where $\chi_{bJ}$ denotes $\chi_{bJ}(1P)$ throughout. This study established the decay mode $\Upsilon(10753)\to\omega\chi_{bJ}$.
Searches at $\sqrt{s}$ = 10.746 GeV for $e^+e^-\to\omega\eta_b(1S)$ and $e^+e^-\to\omega\chi_{b0}$ found no evidence for these processes and set upper limits on the  Born cross sections of 2.5 pb and 7.8 pb at the 90\% confidence level (C.L.) ~\cite{072013}.
An updated measurement of $e^+ e^-\to \Upsilon(nS)\pi^+\pi^-$ ($n$ = 1, 2, 3) improved the precision on the mass, $(10756.6\pm 2.8)$ MeV/$c^2$, and width, $(29.0 \pm 8.9)$ MeV, of the $\Upsilon(10753)$~\cite{Belle-II:2024mjm}.

The radiative decay of the $\Upsilon(10753)$ has not been studied experimentally. Theoretical calculations predict that if the $\Upsilon(10753)$ is a pure $2D$ state, the branching fractions for $\Upsilon(10753)\to\gamma\chi_{b1}$ and $\gamma\chi_{b2}$ are larger than $10^{-2}$~\cite{Godfrey:2015dia,Wang:2018rjg}.
In addition, the cross section distribution for $e^+e^-\to\pi^+\pi^-\Upsilon(nS)$~\cite{220} exhibits two closely spaced peaks corresponding to the $\Upsilon(10753)$ and $\Upsilon(10860)$, analogous to the two-peak structure observed in the cross sections for $e^+e^-\to\pi^+\pi^-J/\psi$ at $Y(4230)$ and $Y(4320)$ by BESIII~\cite{092001}. Evidence for $e^+e^-\to\gamma\chi_{c1}$ and $e^+e^-\to\gamma\chi_{c2}$ was observed by BESIII with statistical significances of 3.0$\sigma$ and 3.4$\sigma$, respectively, using combined data samples at $\sqrt{s}$ = 4.009, 4.230, 4.260, and 4.360 GeV~\cite{BESIII:2014uzr}. If $Y(4230)$ and $\Upsilon(10753)$ share similar internal dynamics, we can expect decays of $e^+e^-\to\gamma\chi_{b1}$ and $e^+e^-\to\gamma\chi_{b2}$ near $\sqrt{s}$ = 10.746 GeV.

We report on a search for the processes $e^+ e^- \to \gamma \chi_{bJ}$ ($J$ = 0, 1, 2) based on electron-positron collisions produced in November 2021 by the SuperKEKB collider~\cite{Akai:2018mbz} at $\sqrt{s}$ = 10.653, 10.701, 10.746, and 10.804 GeV. The data was taken with the Belle II detector~\cite{Belle-II:2010dht}, corresponding to 3.5, 1.6, 9.8, and 4.7 fb$^{-1}$ of integrated luminosity~\cite{2022-056}, respectively. 

The Belle II detector has a cylindrical geometry, with the $z$ axis roughly coinciding with the direction of the electron beam, which defines the forward direction. 
Belle II consists of a two-layer silicon-pixel detector (PXD) surrounded by a four-layer double-sided silicon-strip detector and a 56-layer central drift chamber (CDC). 
These detectors are used to reconstruct the trajectories of charged particles (tracks). Only one-sixth of the second layer of the PXD was installed for the data analyzed here. 
Surrounding the CDC is a time-of-propagation counter (TOP) in the central region, and an aerogel-based ring-imaging Cherenkov counter (ARICH) in the forward region. These detectors, as well as the CDC, provide charged-particle identification. Surrounding TOP and ARICH is an electromagnetic calorimeter (ECL) made of CsI (Tl) crystals, which provides energy and time measurements for photons and electrons. These subsystems are surrounded by a superconducting solenoid, providing an axial magnetic field of 1.5 T. An iron flux return located outside the coil is instrumented with resistive plate chambers and plastic scintillators to detect $K^{0}_{L}$ mesons and to identify muons.

We generate Monte Carlo (MC) simulated signal events for $e^+ e^- \to \gamma \chi_{bJ}$ with {\sc evtgen}~\cite{152} according to a uniform distribution in phase space. Initial-state radiation (ISR) at next-to-leading order accuracy is simulated with {\sc phokhara}~\cite{71}. 
The {\sc geant4}~\cite{250} package is used to simulate the passage of the particles inside the detector and its response. All the data and simulated events are reconstructed and analyzed using the Belle II analysis software~\cite{Kuhr:2018lps}.

The analysis procedure was established and finalized before examining the signal variable distribution in data.
The $\chi_{bJ}$ is reconstructed in the $\gamma\Upsilon(1S)$ final state, with the $\Upsilon(1S)$ decaying to $e^+ e^-$ or $\mu^+ \mu^-$, so the final state particles are $\gamma\gamma l^+ l^-$ ($l$ = $e$ or $\mu$). In the offline analysis, we require the transverse and longitudinal projections of the distances of the closest approach of tracks to the interaction point to be smaller than 1 cm and 3 cm, respectively. We require the number of charged tracks to be two. Charged particle-identification information from various subdetectors is combined in a likelihood ${\cal{L}}_{i}$ for particle species $i$. 
Likelihood ratio ${\cal{R}}_{i}$ = ${\cal{L}}_{i}/{\cal{L}}_{tot}$ are used to identify the species, where ${\cal{L}}_{tot}$ is the sum of likelihoods from electrons, muons, pions, kaons, deuterons and protons.
At least one of the leptons used to reconstruct the $\Upsilon(1S)$ is required to have ${\cal{R}}_{e}$ $>$ 0.9 or ${\cal{R}}_{\mu}$ $>$ 0.9, corresponding to identification efficiencies for electrons and muons of approximately 95\% and 90\%, respectively. Leptons that meet the selection criteria of ${\cal{R}}_{e}$ $>$ 0.9 or ${\cal{R}}_{\mu}$ $>$ 0.9 are considered as electron or muon candidates, respectively. Energy deposits in adjacent electromagnetic calorimeter crystals are treated as photon candidates if they are not associated with charged particles.
To reduce the effects of bremsstrahlung and final-state radiation, photons with energy less than 1 GeV detected in the ECL within a 50 mrad cone of the initial electron or positron direction are included in the calculation of the particle four momentum for the $e^+ e^-$ final states. 

The selection criteria below are optimized by maximizing the figure of merit, defined as $\varepsilon/({\frac{3}{2}+\sqrt{B}})$~\cite{punzi}, where $\varepsilon$ is the signal efficiency determined using simulation, and $B$ is the number of background events estimated in the $\chi_{bJ}$ signal region (9.84 GeV/$c^2$ $<$ $M(\gamma\Upsilon(1S))$ $<$ 9.94 GeV/$c^2$) using simulated background samples, which include the $e^+e^-\to (\gamma)e^+e^-$ (radiative Bhabha scattering) and $e^+e^-\to (\gamma)\mu^+\mu^-$ (radiative di-muon process), scaled to the integrated luminosity of the data.

There are two photons in the $e^+ e^- \to \gamma \chi_{bJ}$ final state. We require the ratio of the energy deposit in a 3 $\times$ 3 matrix of crystals of the ECL to that in the enclosing 5 $\times$ 5 matrix in which the four corner crystals are excluded to be greater than 0.8 for all photons. 
Based on MC simulations, the lower energy photon, denoted $\gamma_{l}$, is assumed to come from the $\chi_{bJ}$, and the higher energy photon, denoted $\gamma_{h}$, from the primary interaction.
The energy for $\gamma_{l}$ is required to be larger than 280 MeV and 290 MeV for $\mu^+ \mu^-$ and $e^+ e^-$ final states, respectively. This different energy requirement is due to different background levels for the $\mu^+ \mu^-$ and $e^+ e^-$ modes. The energy of $\gamma_{h}$ is highly correlated with the invariant mass of $\gamma\Upsilon(1S)$, thus, we do not add any further requirement for its energy. To suppress the Bhabha background, we require the absolute value of the cosine of the  $\gamma_{h}$ polar angle in the $e^+e^-$ center-of-mass frame to be less than 0.7 for the $e^+ e^-$ final state, which results in the largest difference in the reconstruction efficiencies for $\mu^+\mu^-$ and $e^+e^-$ modes.


To reduce background and improve the mass resolution, we perform a kinematic fit for the $\gamma\gamma l^+ l^-$ final state, where the four-momentum and vertex position of the final-state are constrained to the initial $e^+e^-$ collision four-momentum and the interaction region, respectively. In addition, the invariant mass of the dilepton pair is required to lie within the $\Upsilon(1S)$ signal region: $9.44~\mathrm{GeV}/c^2 < M(l^+ l^-) < 9.49~\mathrm{GeV}/c^2$, corresponding to an interval of about $\pm 2.0\sigma$ around the $\Upsilon(1S)$ mass.
Only candidates with $\chi^2 < 30$ from the kinematic fit are retained for further analysis. If multiple candidates are found in an event, the one with the smallest fit $\chi^2$ is selected.
In the data, 2.2\% of selected events contain multiple candidates. Studies based on MC-simulated signal events indicate that no events contain multiple candidates, and the probability of selecting an incorrect signal candidate is negligible.



After imposing all selection criteria, the remaining backgrounds are from di-muon events and Bhabha events, neither of which is expected to peak in the $M(\gamma\Upsilon(1S))$ distribution.

\begin{figure}[htbp]
\centering
\includegraphics[width=8cm]{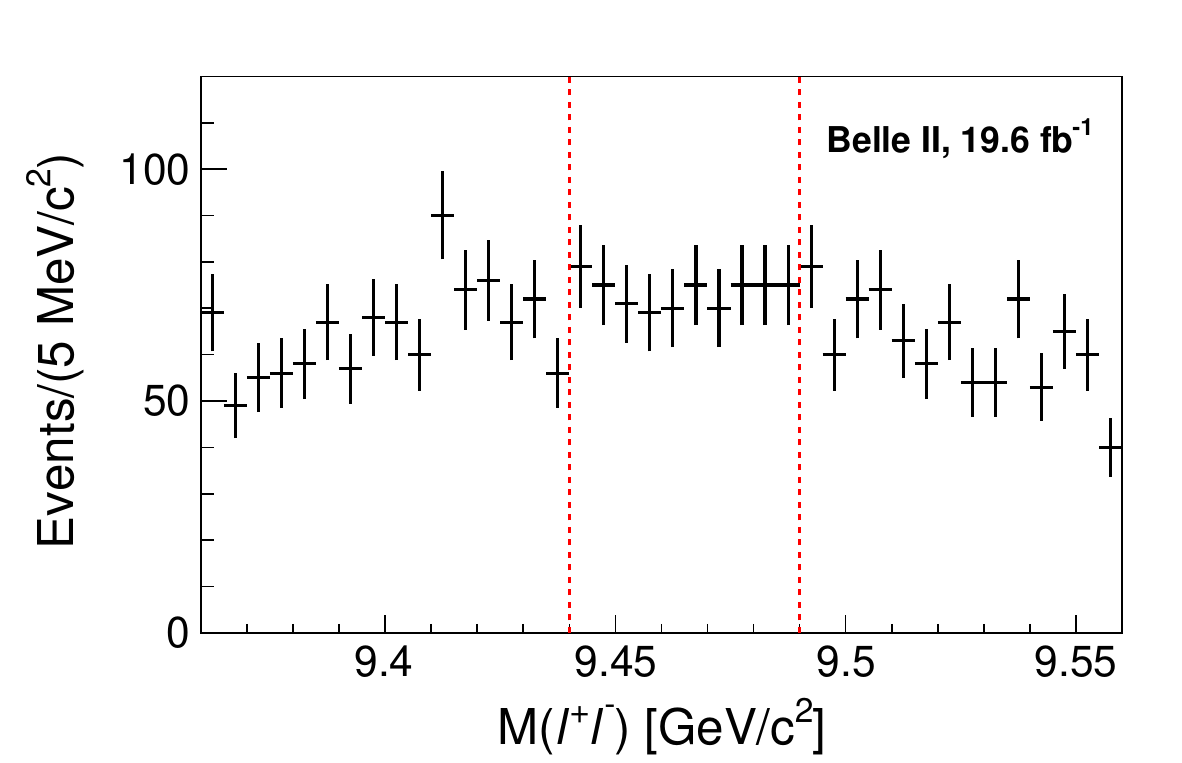}
\caption{The invariant mass spectrum of $l^+l^-$ from a combined $\sqrt{s} = $ 10,653, 10.701, 10.746, and 10.804 GeV data sample. The red dashed lines show the signal region (9.44 GeV/$c^2$ $<$ $M(l^+l^-)$ $<$ 9.49 GeV/$c^2$).}\label{mll}
\end{figure}

\begin{figure*}[htbp]
\centering
\includegraphics[width=6cm]{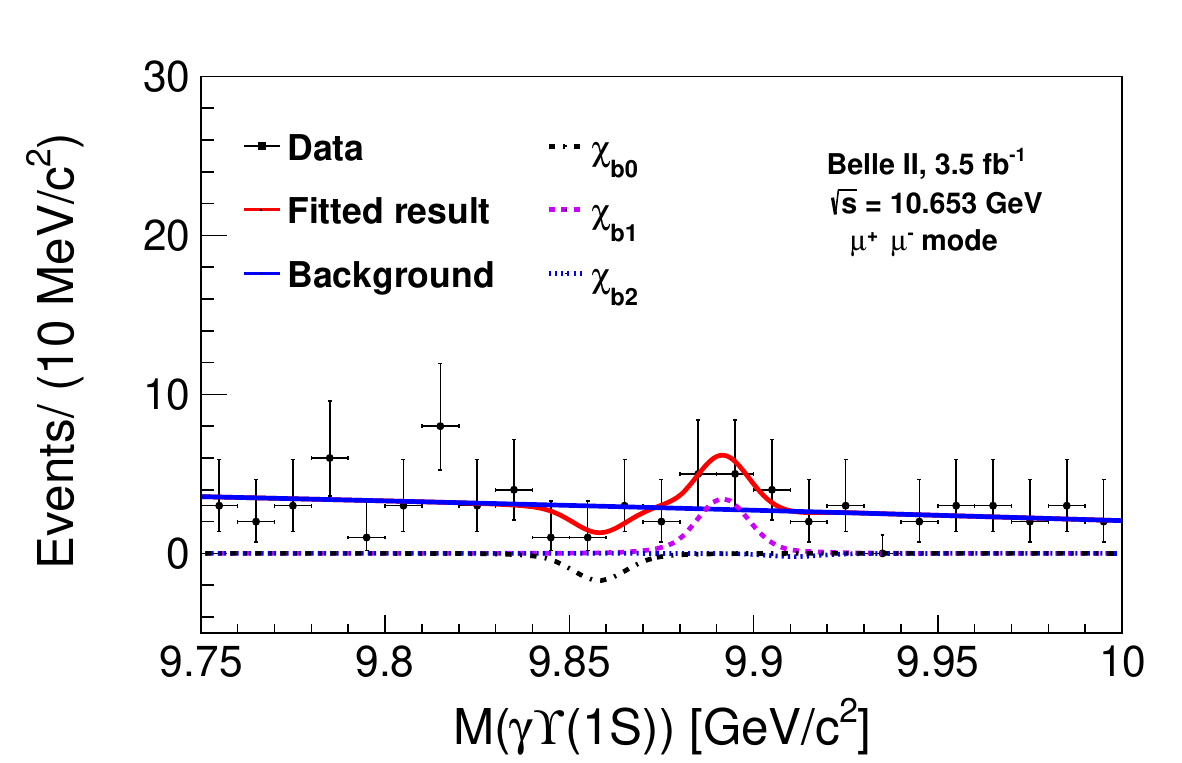}
\put(-170, 100){\bf (a)}
\includegraphics[width=6cm]{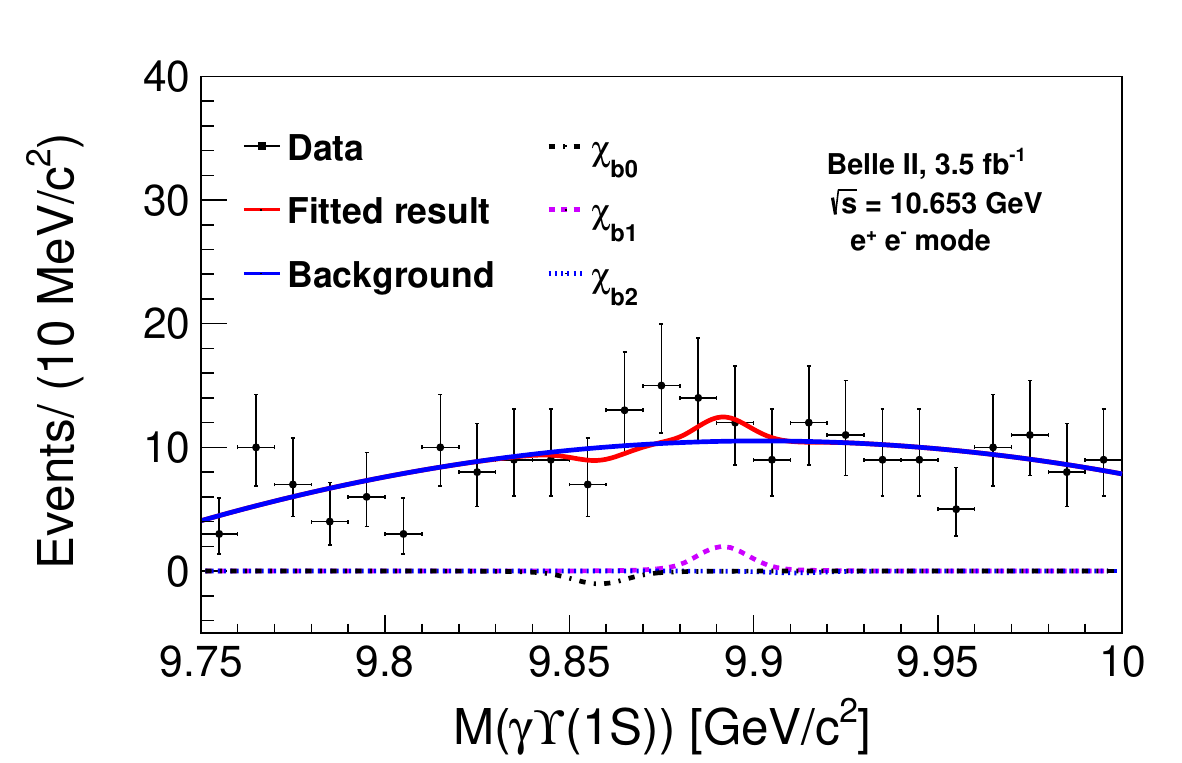}
\put(-170, 100){\bf (b)}

\includegraphics[width=6cm]{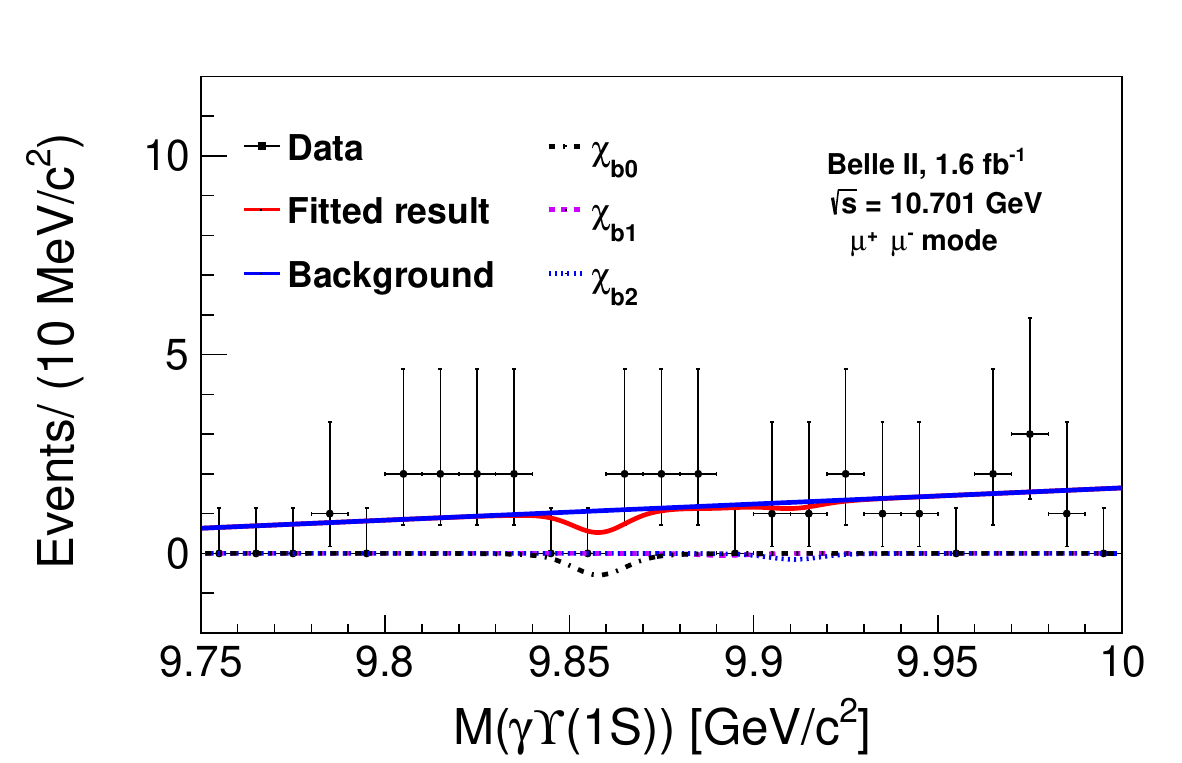}
\put(-170, 100){\bf (c)}
\includegraphics[width=6cm]{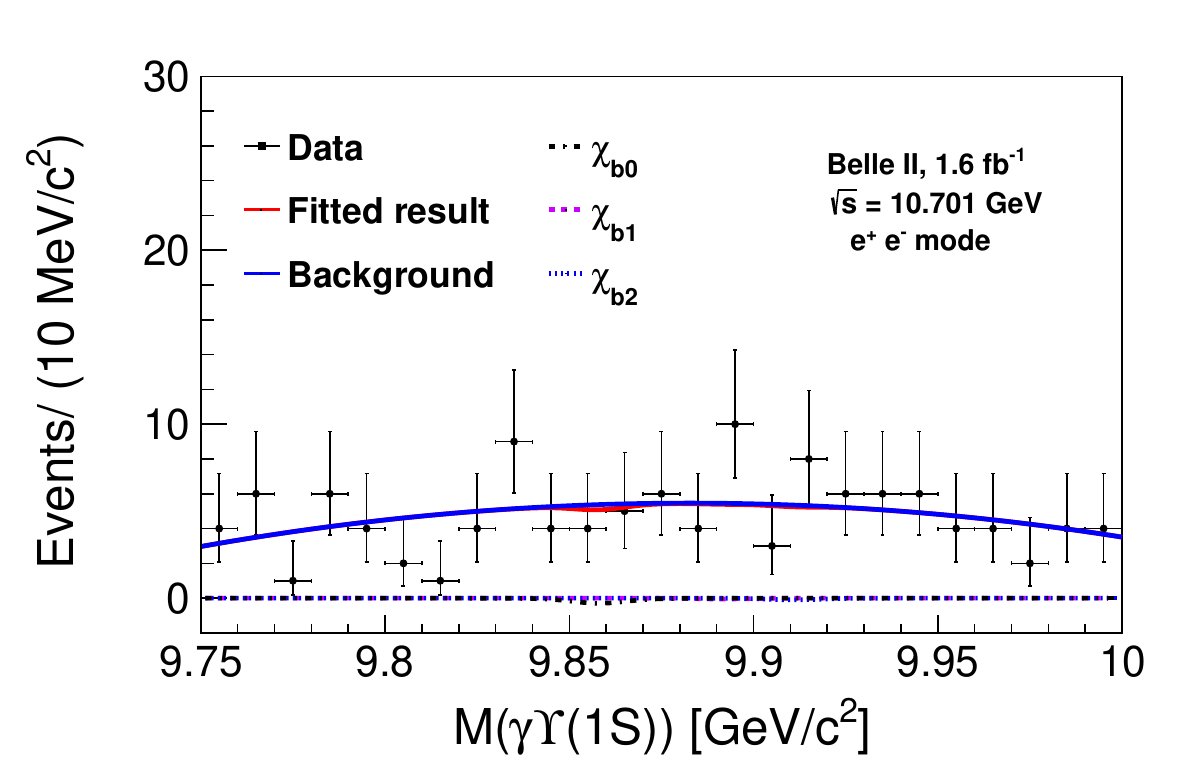}
\put(-170, 100){\bf (d)}

\includegraphics[width=6cm]{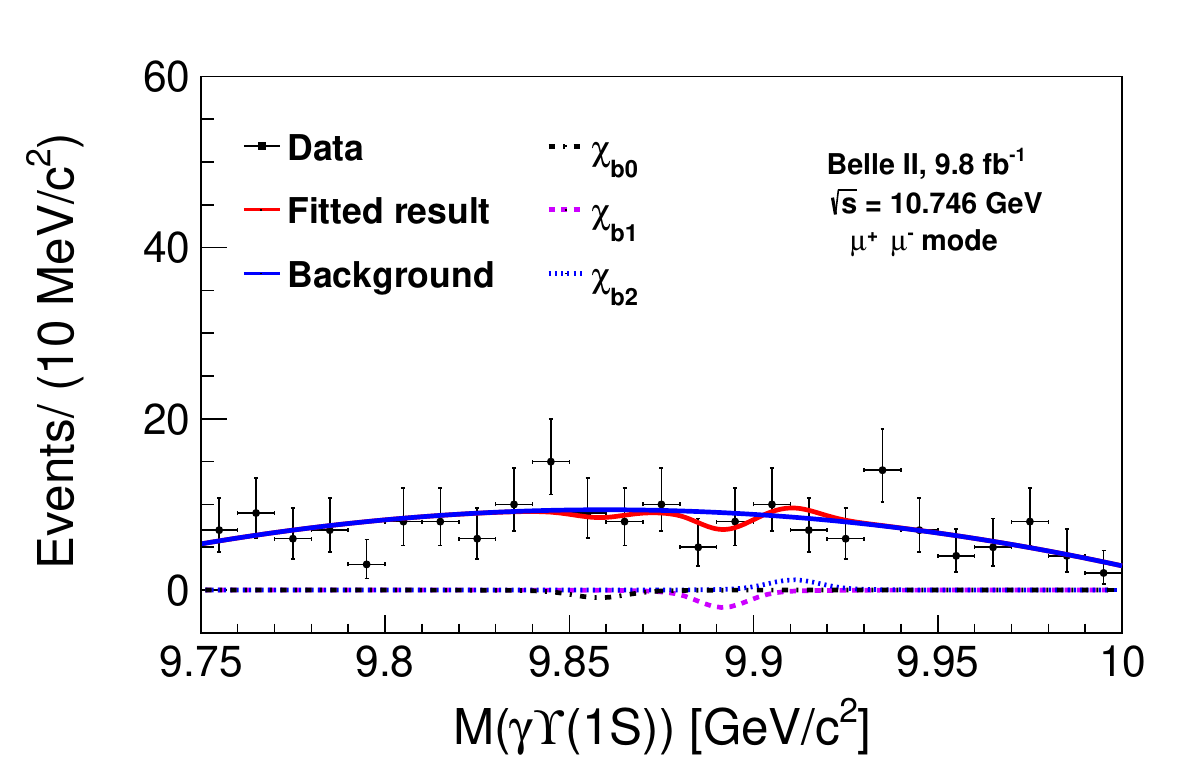}
\put(-170, 100){\bf (e)}
\includegraphics[width=6cm]{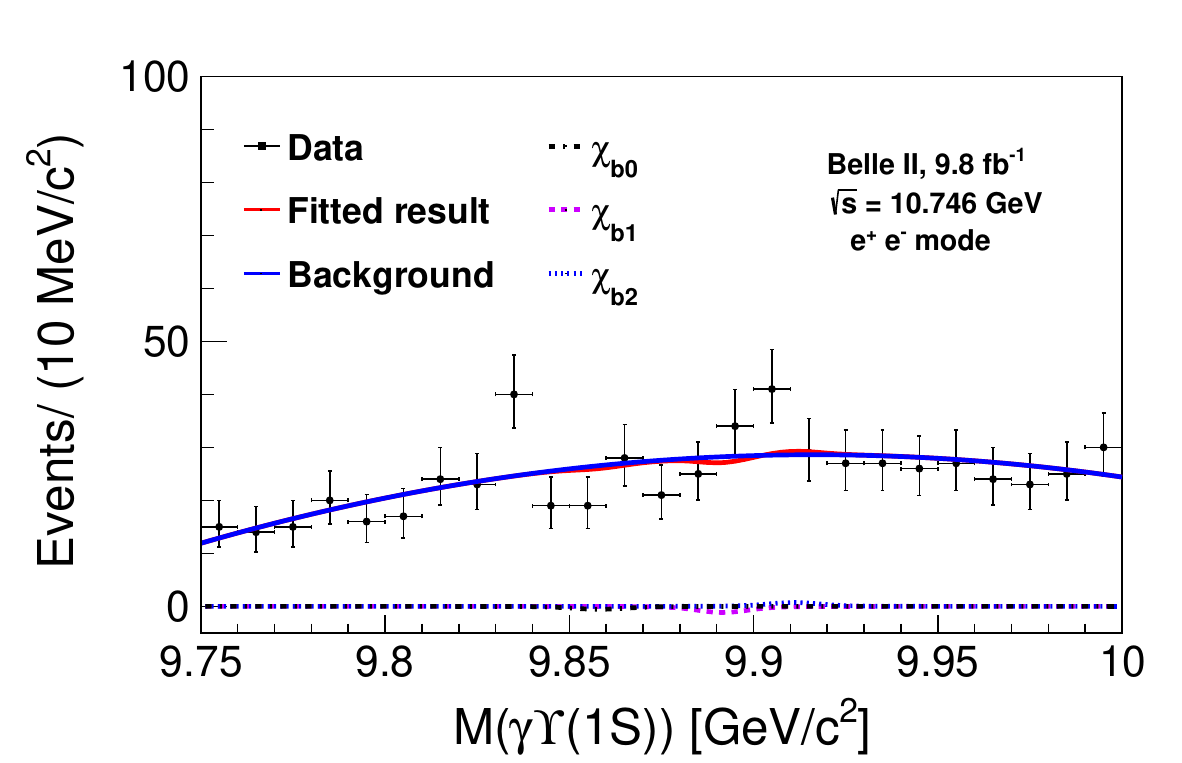}
\put(-170, 100){\bf (f)}

\includegraphics[width=6cm]{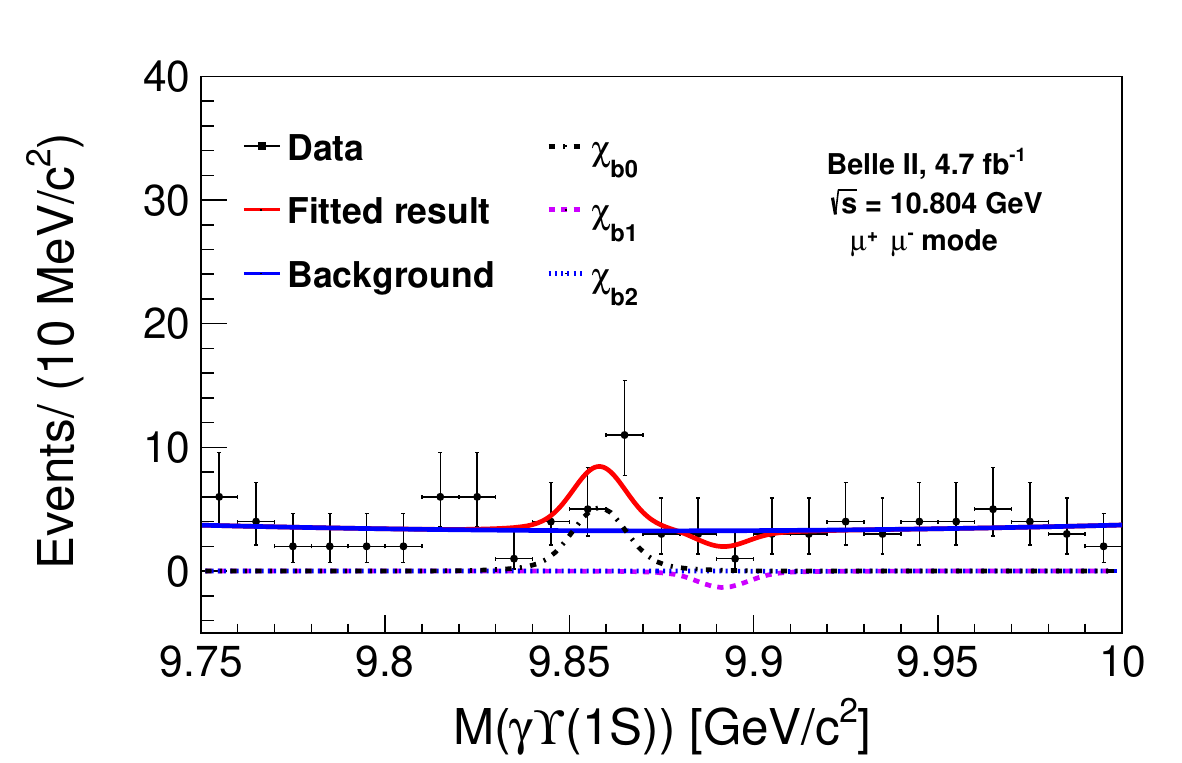}
\put(-170, 100){\bf (g)}
\includegraphics[width=6cm]{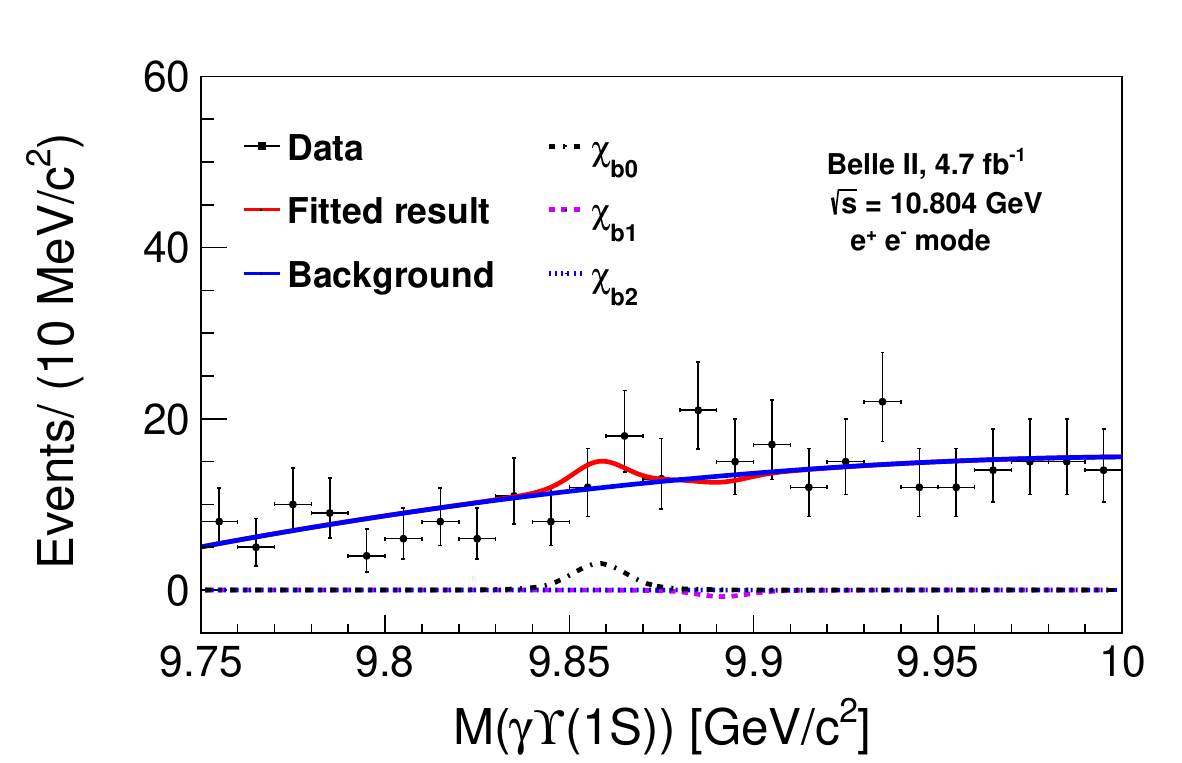}
\put(-170, 100){\bf (h)}
\caption{The fitted results to $M(\gamma\Upsilon(1S))$ distributions from data samples in (left) $\mu^+\mu^-$ and (right) $e^+e^-$ modes at $\sqrt{s}$ $=$ (a, b) 10.653, (c, d) 10.701, (e, f) 10.746, and (g, h) 10.804 GeV, respectively. Here, the points with error bars are the data samples. The red curves show the fitted results. The blue curves show the total backgrounds. The black dot-dashed curves show the $\chi_{b0}$ candidate, the purple dashed curves show the $\chi_{b1}$ candidate, and the blue dotted curves show the $\chi_{b2}$ candidate.}
\label{simultaneous_fit}
\end{figure*}

With the application of the above requirements except for the $\Upsilon(1S)$ signal region selection, the $M(l^+l^-)$ distribution from the combined $\sqrt{s} = $ 10.653, 10.701, 10.746, and 10.804 GeV data sample is shown in Fig.~\ref{mll}. The red dashed lines in Fig.~\ref{mll} show the $\Upsilon(1S)$ signal region. No clear $\Upsilon(1S)$ signal is observed. The $\gamma\Upsilon(1S)$ invariant mass distributions are shown in Fig.~\ref{simultaneous_fit} for candidates in the $\Upsilon(1S)$ signal region.

We perform unbinned extended maximum-likelihood fits to $\gamma\Upsilon(1S)$ invariant mass distributions to extract signal yields. The fitted results are shown in Fig.~\ref{simultaneous_fit}.
We simultaneously fit the $\Upsilon(1S)\to\mu^+\mu^-$ and $\Upsilon(1S)\to e^+ e^-$ modes for each energy point. In the simultaneous fit, the ratio of yields between the $\mu^+\mu^-$ and $e^+e^-$ channels is fixed according to the product of their respective branching fractions~\cite{PDG} and reconstruction efficiencies.
The probability density functions (PDF) of $\chi_{bJ}$ signals are modeled as a sum of a Crystal Ball function~\cite{CB} and a Gaussian function that share a common mean. The means are fixed to the known $\chi_{bJ}$ masses~\cite{PDG}, while the remaining parameters are derived from simulation. To account for data-simulation discrepancies, the photon energy in simulated events is smeared according to a resolution scale factor, $1.01\pm 0.01$, obtained from a $\pi^0\to\gamma\gamma$ control sample. A second-order Chebyshev polynomial is used to describe the background.
The signal significances are estimated using $\sqrt{-2\ln(\mathcal{L}_0/\mathcal{L}_{\rm max})}$,
where $\mathcal{L}_0$ and $\mathcal{L}_{\rm max}$ are the maximized likelihoods without and with the signal, respectively~\cite{Wilks:1938dza}.
The signal yields at each energy point are listed in Table~\ref{t1}.
The largest statistical signal significance, $2.3\sigma$, is observed for $\gamma\chi_{b0}$ at $\sqrt{s}=10.804$ GeV.
Since no significant signal was observed, we compute 90\% C.L. upper limits on the signal yields, $N^{\rm UL}$, by solving the equation 
\begin{equation}\label{eq:eq1}
    \int _0^{N^{\rm UL}}{\cal L}(x)dx/\int _0^{+\infty}{\cal L}(x)dx = 0.90,
\end{equation}
where $x$ is the assumed signal yield, and ${\cal L}(x)$ is the corresponding maximized profiled likelihood of the fit. 
To obtain ${\cal L}(x)$, we perform a simultaneous fit to $\gamma\Upsilon(1S)$ invariant mass distributions for each fixed value of $x$, while the yields of other signals and background, as well as the parameters of background shape, are allowed to float.
The upper limits on the signal yields, including systematic uncertainties (discussed below), are listed in Table~\ref{t1}.

The Born cross sections are calculated as
\begin{equation}\label{eq:eq2}
\sigma_{\rm Born}(e^+ e^- \to \gamma \chi_{bJ}) = \frac{N^{\rm sig} |1-\Pi|^2}{\sum\limits_{i = e,\mu} (\varepsilon_{i}\BR^{\rm int}_{i}){\cal L}(1+\delta_{\rm ISR})},
\end{equation}
where $N^{\rm sig}$ is the signal yield combined from $\mu^+\mu^-$ and $e^+e^-$ modes,
${\cal L}$ is the integrated luminosity of the data sample, 
$\varepsilon_{i}$ and $\BR^{\rm int}_{i}$ are the reconstruction efficiency and product branching fraction of the intermediate states for the $\Upsilon(1S)\to e^+e^-$ or $\Upsilon(1S)\to\mu^+\mu^-$ modes.
The factor $|1-\Pi|$ is the vacuum polarization factor~\cite{585}. 
The radiative correction factor~\cite{isr,113009,2605}, $(1+\delta_{\rm ISR})$, is calculated with the assumption that the $e^+ e^- \to \gamma \chi_{bJ}$ cross section follows the $1/s$ line shape.
The 90\% C.L. upper limits on the Born cross sections are calculated using Eq.(\ref{eq:eq1}), with $x\equiv\sigma_{\rm Born}$ as defined in Eq.(\ref{eq:eq2}).
The upper limits on the Born cross sections including systematic uncertainties (discussed below) for $e^+ e^-\to\gamma\chi_{bJ}$ at $\sqrt{s}$ = 10.653, 10.701, 10.746, and 10.804 GeV are listed in Table~\ref{t1}.

\begin{table*}[htbp!]
\caption{The values used to determine the Born cross sections for $e^+ e^- \to \gamma \chi_{bJ}$ ($J$ = 0, 1, 2) at $\sqrt{s}$ = 10.653, 10.701, 10.746, and 10.804 GeV, respectively. Here ${\cal L}$ is the integrated luminosity, $N^{\rm fit}$ is the sum of signal yields of $\Upsilon(1S)\to e^+e^-$ and $\Upsilon(1S)\to \mu^+\mu^-$ modes with purely statistical uncertainty, $N^{\rm UL}$ is the upper limit at 90\% C.L. on the signal yield, $\varepsilon_{e^+e^-}$ and $\varepsilon_{\mu^+\mu^-}$ are reconstruction efficiencies for $\Upsilon(1S)\to e^+e^-$ and $\Upsilon(1S)\to\mu^+\mu^-$ modes, $\BR^{\rm int}_{e^+e^-}$ and $\BR^{\rm int}_{\mu^+\mu^-}$ are the product of the branching fractions of the intermediate states for $\Upsilon(1S)\to e^+e^-$ and $\Upsilon(1S)\to\mu^+\mu^-$ modes, $|1-\Pi|^2$ is the vacuum polarization factor, $1+\delta_{\rm ISR}$ is the radiative correction factor, syst is the systematic uncertainty, and $\sigma^{\rm UL}_{\rm Born}$ is the upper limit at 90\% C.L. on the Born cross section.}\label{t1}
\vspace{0.2cm}
\centering
\begin{tabular}{c c c r r c c c c c c r r}
\hline\hline
$\sqrt{s}$ (GeV) & ${\cal L}$ (fb$^{-1}$) & Channel & \multicolumn{1}{c}{$N^{\rm fit}$} & \multicolumn{1}{c}{$N^{\rm UL}$} & $\varepsilon_{e^+e^-}$ & $\varepsilon_{\mu^+\mu^-}$ & $\BR^{\rm int}_{e^+e^-}$ & $\BR^{\rm int}_{\mu^+\mu^-}$ & $|1-\Pi|^2$ & $1+\delta_{\rm ISR}$ & \multicolumn{1}{c}{syst (\%)} & \multicolumn{1}{c}{$\sigma^{\rm UL}_{\rm Born}$ (pb)} \\\hline
& & $\gamma\chi_{b0}$ & $-$5.5 $\pm$ 4.9 & 8.5 & 0.215 & 0.361 & 0.00046 & 0.00048 & 0.930 & 0.901 & 15.5 & 9.18\\
10.653 & 3.512 & $\gamma\chi_{b1}$ & 10.4 $\pm$ 7.0 & 21.5 & 0.223 & 0.353 & 0.00841 & 0.00873 & 0.930 & 0.898&  9.0& 1.28\\
& & $\gamma\chi_{b2}$ & $-$0.7 $\pm$ 5.5 & 10.5 & 0.212 & 0.341 & 0.00430 & 0.00446
 & 0.930 & 0.896 &  8.9 & 1.28\\\hline
& &$\gamma\chi_{b0}$ & $-$1.7 $\pm$ 2.8 & 7.0 & 0.214 & 0.346 & 0.00046 & 0.00048 & 0.931 & 0.905&  15.5& 16.68 \\
10.701 & 1.632 & $\gamma\chi_{b1}$ & $-$0.2 $\pm$ 3.7 & 8.6 & 0.217 & 0.351 & 0.00841 & 0.00873 & 0.931 & 0.903& 9.0 & 1.11\\
& & $\gamma\chi_{b2}$ & $-$0.5 $\pm$ 3.4 &  8.0 & 0.219 & 0.341 & 0.00430 & 0.00446 & 0.931 & 0.901&  8.9& 2.06 \\\hline
& &$\gamma\chi_{b0}$ & $-$2.9 $\pm$ 9.5 &  13.9 & 0.216 & 0.356 & 0.00046 & 0.00048 & 0.931 & 0.909 & 15.5 & 5.37 \\
10.746 & 9.818 & $\gamma\chi_{b1}$ & $-$6.1 $\pm$ 8.5 & 12.1 & 0.212 & 0.361 & 0.00841 & 0.00873 & 0.931 & 0.906&  9.0& 0.26\\
& & $\gamma\chi_{b2}$ & 3.8 $\pm$ 9.7 & 19.1 & 0.218 & 0.358 & 0.00430 & 0.00446 & 0.931 & 0.904&  8.9& 0.79\\\hline
& &$\gamma\chi_{b0}$ & 16.7 $\pm$ 8.1 & 30.1 & 0.222 & 0.361 & 0.00046 & 0.00048 & 0.932 & 0.914& 15.5 & 23.77 \\
10.804 & 4.689 & $\gamma\chi_{b1}$ & $-$4.0 $\pm$ 5.6 & 10.0 & 0.224 & 0.366 & 0.00841 & 0.00873 & 0.932 & 0.911& 9.0 & 0.43\\
& & $\gamma\chi_{b2}$ & $-$0.1 $\pm$ 5.7 & 11.2 & 0.228 & 0.364 & 0.00430 & 0.00446 & 0.932 & 0.909 & 8.9 & 0.94\\\hline\hline
\end{tabular}
\end{table*}

The systematic uncertainties in the cross-section measurements of $e^+ e^- \to \gamma \chi_{bJ}$ include contributions from the branching fractions of intermediate states, reconstruction efficiency uncertainties, trigger simulation, MC statistical uncertainty, integrated luminosity,  angular distributions, beam-energy calibration, the radiative correction factor, the masses and widths of $\chi_{bJ}$, and the fit model. The systematic uncertainties from the masses and widths of $\chi_{bJ}$ and the fit model are additive. The other sources of systematic uncertainties are multiplicative, and are listed in Table~\ref{t2}.

The main source of the systematic uncertainty comes from the uncertainties of branching fractions of intermediate states, which are taken from the Review of Particle Physics~\cite{PDG}. The total uncertainties from intermediate branching fractions in $e^+e^-\to\gamma\chi_{b0}$, $e^+e^-\to\gamma\chi_{b1}$, and $e^+e^-\to\gamma\chi_{b2}$ are 14.6\%, 7.3\%, and 7.2\%, respectively.

For the lepton track reconstruction efficiency, an uncertainty of 0.3\% per track, evaluated on $e^+e^-\to\tau^+\tau^-$ events, is included in the systematic uncertainty.
In our event selection, we only require at least one lepton to be identified. Consequently, the lepton identification efficiency is so high ( $>$ 99\%) that the corresponding systematic uncertainty is negligible.
The systematic uncertainty of the photon reconstruction efficiency is 1.1\% for $\gamma_{l}$ and 0.7\% for $\gamma_{h}$, derived using the $e^+e^-\to (\gamma) \mu^+\mu^-$ sample.
We add individual uncertainties in quadrature to obtain the final uncertainties related to the reconstruction efficiency, which is 1.9\%.

The simulated trigger efficiency differs from the efficiency measured on data, as studied on Bhabha and dimuon control samples. We require the Bhabha and dimuon samples to satisfy the same selection criteria as the signal process $e^+e^-\to\gamma\chi_{bJ}$, as well as hardware-based level 1 trigger requirements in the CDC and ECL~\cite{2022-056}. The efficiencies from data and simulation are 99.8\% and 99.6\% for the dimuon process, and 98.6\% and 96.5\% for the Bhabha process. As a result, we include 0.2\% uncertainty for the $\mu^+\mu^-$ mode and 2.1\% for the $e^+ e^-$ mode. We combine these uncertainties, taking into account the detection efficiencies for $\mu^+\mu^-$ and $e^+e^-$ modes as weight factors, to obtain 1.7\%.

The statistical uncertainty in the MC simulation efficiency is 1\%.
The systematic uncertainty of the luminosity is estimated to be 1.2\%, using Bhabha, di-photon, and di-muon event samples~\cite{2022-056}.

We change the assumed angular distribution from uniform to $1\pm {\rm cos}^2\theta$ for $\gamma\chi_{bJ}$, where $\theta$ is the polar angle of $\gamma$ or $\chi_{bJ}$ in $e^+ e^-$ center-of-mass frame, and take the maximum difference in efficiency (4$\%$) as the uncertainty. 

The uncertainties of the collision energies are $\pm$1.2 MeV, $\pm$0.8 MeV, $\pm$0.7 MeV, and $\pm$0.85 MeV for center-of-mass energies at 10.653, 10.701, 10.746, and 10.804 GeV respectively, according to the measurement of the energy dependence of the $e^+e^-\to B\bar B$, $B\bar B^*$, and $B^*\bar B^*$ cross sections at Belle II~\cite{2022-015}. We vary the collision energies in the kinematic fit to the MC simulated signal candidates according to uncertainties of the collision energies above, and take the largest difference in calculated signal reconstruction efficiency as the uncertainty due to beam-energy calibration, which is 1\%. We also account for the 0.3\% uncertainty associated with the position of the center of the interaction region in the calculation of the signal reconstruction efficiency. This effect is found to be negligible.

To calculate the radiative correction factor for $e^+e^-\to\gamma\chi_{bJ}$, we assume the dressed cross section falls as $1/s$.
We change the $1/s$ to $1/s^2$~\cite{092015} and take the difference in the radiative correction factor as the systematic uncertainty. The uncertainties on the radiative correction factor in $e^+e^-\to\gamma\chi_{b0}$, $e^+e^-\to\gamma\chi_{b1}$, and $e^+e^-\to\gamma\chi_{b2}$ are 1.5\%, 1.4\%, and 1.4\%, respectively.


All the multiplicative uncertainties in the measurements of $\sigma_{\rm Born}(e^+e^-\to\gamma\chi_{bJ})$ are added in quadrature to obtain the total systematic uncertainty.


The additive systematic uncertainties are evaluated by varying input assumptions and parameters and assigning the differences in cross section as systematic uncertainties.
We considered the following systematic uncertainties associated with the masses and widths of $\chi_{bJ}$. We varied the mean of the signal PDF according to the uncertainties of the $\chi_{bJ}$ masses from Review of Particle Physics ~\cite{PDG}, which are $\pm0.42$~MeV, $\pm0.26$~MeV, and $\pm0.26$~MeV for $\chi_{b0}$, $\chi_{b1}$, and $\chi_{b2}$, respectively. To account for possible differences in mass resolution between data and MC simulation, dominated by the photon energy resolution, we repeated the fit while varying the photon energy resolution scale factor ($1.01\pm0.01$) within its uncertainty by $\pm 1\sigma$.

The systematic uncertainties associated with the fit model are estimated by changing the order of the background polynomial and the range of the fit with all the $\chi_{b0}$, $\chi_{b1}$, and $\chi_{b2}$ components included. Due to the small branching fraction for $\chi_{b0}\to\gamma\Upsilon(1S)$, we compare the fit results without a $\chi_{b0}$ component for the $e^+e^-\to\gamma\chi_{b1}$ and $e^+e^-\to\gamma\chi_{b2}$ processes. 
The absolute uncertainties of branching fractions for $\Upsilon(1S)\to\mu^+\mu^-$ and $\Upsilon(1S)\to e^+ e^-$ are $\pm 0.04\%$ and $\pm 0.08\%$, respectively~\cite{PDG}, which affect the ratio of yields between the $\mu^+\mu^-$ and $e^+e^-$ modes in the simultaneous fit to $\Upsilon(1S)\to\mu^+\mu^-$ and $\Upsilon(1S)\to e^+ e^-$.
Therefore, in the simultaneous fit, we change their branching fractions within uncertainties.

Systematic uncertainties are incorporated into our calculations of upper limits in two steps. First, in determining the additive systematic uncertainty from the masses and widths of $\chi_{bJ}$ and fit model (as described above), we additionally calculate the upper limit for each possible fit configuration and determine the most conservative upper limit at 90\% C.L. on the number of signal events. Then, the likelihood with that most conservative upper limit is convolved with a Gaussian function whose width is equal to the corresponding total multiplicative uncertainty summarized in Table~\ref{t2}. The resulting 90\% C.L. upper limits on the Born cross section are tabulated in the rightmost column of Table~\ref{t1}.

\begin{table*}[htbp!]
\centering
\caption{Relative multiplicative systematic uncertainties (\%) of the measurements of the Born cross sections for $e^+ e^- \to \gamma \chi_{b0}$, $e^+ e^- \to \gamma \chi_{b1}$, and $e^+ e^- \to \gamma \chi_{b2}$ at $\sqrt{s}$ near 10.746$~\gev$.}\label{t2}
\begin{tabular}{c r c c }
\hline\hline
Sources & \multicolumn{1}{c}{$\gamma\chi_{b0}$} & $\gamma\chi_{b1}$ & $\gamma\chi_{b2}$\\
\hline
Branching fractions &14.6 &7.3 &7.2 \\
Detection efficiency  &1.9 &1.9 &1.9\\
Trigger simulation  &1.7 &1.7 &1.7\\
MC sample size &1.0 &1.0 &1.0\\
Integrated luminosity &1.2 &1.2 &1.2\\
Angular distributions &3.9 &3.9 &4.0\\
Beam-energy calibration &1.0 &1.0 &1.0\\
Radiative correction factor&1.5 &1.4 &1.4 \\
\hline
Sum & 15.5& 9.0& 8.9\\\hline\hline
\end{tabular}
\end{table*}

In summary, we report a search for the $e^+ e^- \to \gamma \chi_{bJ}$ ($J$ = 0, 1, 2) processes at $\sqrt{s}$ = 10.653, 10.701, 10.746 and 10.804 GeV. 
We do not find evidence for any signal process, and set upper limits at 90\% C.L. on the corresponding Born cross sections.
Compared with hadronic bottomonium transitions such as $\omega\chi_{bJ}$ ($J$ = 1, 2), $\pi^+\pi^-\Upsilon(nS)$ ($n$ = 1, 2), and $\eta \Upsilon(2S)$ ~\cite{Belle-II:2022xdi,Belle-II:2024mjm,Belle-II:2025ubm}, radiative transition cross sections are significantly smaller. Our measurements, when combined with existing results on hadronic transitions, provide valuable input for clarifying the underlying nature of the $\Upsilon(10753)$.
This work, based on data collected using the Belle II detector, which was built and commissioned prior to March 2019,
was supported by
Higher Education and Science Committee of the Republic of Armenia Grant No.~23LCG-1C011;
Australian Research Council and Research Grants
No.~DP200101792, 
No.~DP210101900, 
No.~DP210102831, 
No.~DE220100462, 
No.~LE210100098, 
and
No.~LE230100085; 
Austrian Federal Ministry of Education, Science and Research,
Austrian Science Fund (FWF) Grants
DOI:~10.55776/P34529,
DOI:~10.55776/J4731,
DOI:~10.55776/J4625,
DOI:~10.55776/M3153,
and
DOI:~10.55776/PAT1836324,
and
Horizon 2020 ERC Starting Grant No.~947006 ``InterLeptons'';
Natural Sciences and Engineering Research Council of Canada, Digital Research Alliance of Canada, and Canada Foundation for Innovation;
Fundamental Research Funds of China for the Central Universities No.~2242025RCB0014 and No.~RF1028623046;
National Key R\&D Program of China under Contract No.~2024YFA1610503,
and
No.~2024YFA1610504
National Natural Science Foundation of China and Research Grants
No.~12475076,
No.~11575017,
No.~11761141009,
No.~11705209,
No.~11975076,
No.~12135005,
No.~12150004,
No.~12161141008,
No.~12405099,
No.~12475093,
and
No.~12175041,
and Shandong Provincial Natural Science Foundation Project~ZR2022JQ02;
the Czech Science Foundation Grant No. 22-18469S,  Regional funds of EU/MEYS: OPJAK
FORTE CZ.02.01.01/00/22\_008/0004632 
and
Charles University Grant Agency project No. 246122;
European Research Council, Seventh Framework PIEF-GA-2013-622527,
Horizon 2020 ERC-Advanced Grants No.~267104 and No.~884719,
Horizon 2020 ERC-Consolidator Grant No.~819127,
Horizon 2020 Marie Sklodowska-Curie Grant Agreement No.~700525 ``NIOBE''
and
No.~101026516,
and
Horizon 2020 Marie Sklodowska-Curie RISE project JENNIFER2 Grant Agreement No.~822070 (European grants);
L'Institut National de Physique Nucl\'{e}aire et de Physique des Particules (IN2P3) du CNRS
and
L'Agence Nationale de la Recherche (ANR) under Grant No.~ANR-21-CE31-0009 (France);
BMFTR, DFG, HGF, MPG, and AvH Foundation (Germany);
Department of Atomic Energy under Project Identification No.~RTI 4002,
Department of Science and Technology,
and
UPES SEED funding programs
No.~UPES/R\&D-SEED-INFRA/17052023/01 and
No.~UPES/R\&D-SOE/20062022/06 (India);
Israel Science Foundation Grant No.~2476/17,
U.S.-Israel Binational Science Foundation Grant No.~2016113, and
Israel Ministry of Science Grant No.~3-16543;
Istituto Nazionale di Fisica Nucleare and the Research Grants BELLE2,
and
the ICSC – Centro Nazionale di Ricerca in High Performance Computing, Big Data and Quantum Computing, funded by European Union – NextGenerationEU;
Japan Society for the Promotion of Science, Grant-in-Aid for Scientific Research Grants
No.~16H03968,
No.~16H03993,
No.~16H06492,
No.~16K05323,
No.~17H01133,
No.~17H05405,
No.~18K03621,
No.~18H03710,
No.~18H05226,
No.~19H00682, 
No.~20H05850,
No.~20H05858,
No.~22H00144,
No.~22K14056,
No.~22K21347,
No.~23H05433,
No.~26220706,
and
No.~26400255,
and
the Ministry of Education, Culture, Sports, Science, and Technology (MEXT) of Japan;  
National Research Foundation (NRF) of Korea Grants
No.~2021R1-F1A-1064008, 
No.~2022R1-A2C-1003993,
No.~2022R1-A2C-1092335,
No.~RS-2016-NR017151,
No.~RS-2018-NR031074,
No.~RS-2021-NR060129,
No.~RS-2023-00208693,
No.~RS-2024-00354342
and
No.~RS-2025-02219521,
Radiation Science Research Institute,
Foreign Large-Size Research Facility Application Supporting project,
the Global Science Experimental Data Hub Center, the Korea Institute of Science and
Technology Information (K25L2M2C3 ) 
and
KREONET/GLORIAD;
Universiti Malaya RU grant, Akademi Sains Malaysia, and Ministry of Education Malaysia;
Frontiers of Science Program Contracts
No.~FOINS-296,
No.~CB-221329,
No.~CB-236394,
No.~CB-254409,
and
No.~CB-180023, and SEP-CINVESTAV Research Grant No.~237 (Mexico);
the Polish Ministry of Science and Higher Education and the National Science Center;
the Ministry of Science and Higher Education of the Russian Federation
and
the HSE University Basic Research Program, Moscow;
University of Tabuk Research Grants
No.~S-0256-1438 and No.~S-0280-1439 (Saudi Arabia), and
Researchers Supporting Project number (RSPD2025R873), King Saud University, Riyadh,
Saudi Arabia;
Slovenian Research Agency and Research Grants
No.~J1-50010
and
No.~P1-0135;
Ikerbasque, Basque Foundation for Science,
State Agency for Research of the Spanish Ministry of Science and Innovation through Grant No. PID2022-136510NB-C33, Spain,
Agencia Estatal de Investigacion, Spain
Grant No.~RYC2020-029875-I
and
Generalitat Valenciana, Spain
Grant No.~CIDEGENT/2018/020;
The Knut and Alice Wallenberg Foundation (Sweden), Contracts No.~2021.0174 and No.~2021.0299;
National Science and Technology Council,
and
Ministry of Education (Taiwan);
Thailand Center of Excellence in Physics;
TUBITAK ULAKBIM (Turkey);
National Research Foundation of Ukraine, Project No.~2020.02/0257,
and
Ministry of Education and Science of Ukraine;
the U.S. National Science Foundation and Research Grants
No.~PHY-1913789 
and
No.~PHY-2111604, 
and the U.S. Department of Energy and Research Awards
No.~DE-AC06-76RLO1830, 
No.~DE-SC0007983, 
No.~DE-SC0009824, 
No.~DE-SC0009973, 
No.~DE-SC0010007, 
No.~DE-SC0010073, 
No.~DE-SC0010118, 
No.~DE-SC0010504, 
No.~DE-SC0011784, 
No.~DE-SC0012704, 
No.~DE-SC0019230, 
No.~DE-SC0021274, 
No.~DE-SC0021616, 
No.~DE-SC0022350, 
No.~DE-SC0023470; 
and
the Vietnam Academy of Science and Technology (VAST) under Grants
No.~NVCC.05.02/25-25
and
No.~DL0000.05/26-27.

These acknowledgements are not to be interpreted as an endorsement of any statement made
by any of our institutes, funding agencies, governments, or their representatives.

We thank the SuperKEKB team for delivering high-luminosity collisions;
the KEK cryogenics group for the efficient operation of the detector solenoid magnet and IBBelle on site;
the KEK Computer Research Center for on-site computing support; the NII for SINET6 network support;
and the raw-data centers hosted by BNL, DESY, GridKa, IN2P3, INFN, 
and the University of Victoria.

\renewcommand{\baselinestretch}{1.2}

\end{document}